\documentclass[aip,apl,reprint,groupedaddress,nofootinbib]{revtex4-1}

\usepackage{graphicx}
\usepackage{color}
\definecolor{grey}{gray}{.8}

\usepackage{upgreek}
\usepackage{xspace}

\newcommand{\fig}[1]{Fig.$\;$#1}
\newcommand{\um}{$\upmu$m\xspace}
\newcommand{\uL}{$\upmu$L\xspace}

\draft % marks overfull lines with a black rule on the right

\begin{document}

%\preprint{}

\title{Refractometric sensing of Li salt with visible-light Si$_3$N$_4$ microdisk resonators} %Title of paper


\author{C. Doolin}
\email[]{doolin@ualberta.ca}
\author{P. Doolin}
\author{B.C. Lewis}
\author{J.P. Davis}
\email[]{jdavis@ualberta.ca}
\affiliation{Department of Physics, University of Alberta, T6G 2E1 Edmonton, AB, Canada}

\date{\today}

\begin{abstract}

We demonstrate aqueous refractive index sensing with 15 to 30 \um diameter silicon nitride microdisk resonators to detect small concentrations of Li salt.  A dimpled-tapered fiber is used to couple 780 nm visible light to the microdisks, in order to perform spectroscopy their optical resonances.  The dimpled fiber probe allows testing of multiple devices on a chip in a single experiment.  This sensing system is versatile and easy to use, while remaining competitive with other refractometric sensors.  For example, from a 20 \um diameter device we measure a sensitivity of $200 \pm 30$ nm/RIU with a loaded quality factor of $1.5 \times 10^{4}$, and a limit of detection down to $(1.3 \pm 0.1)\times10^{-6}$ RIU. 
\end{abstract}

%\pacs{}% insert suggested PACS numbers in braces on next line
\maketitle %\maketitle must follow title, authors, abstract and \pacs

Optical whispering-gallery mode (WGM) resonators are an area under avid research as they promise fast, sensitive and label-free detection of chemical and biological samples.\cite{Luchansky11, Vollmer12, Meldrum14}  Sensors based on whispering-gallery mode resonators have been used for the label-free detection of single viruses,\cite{Volmer08, McClellan12} nanoparticles,\cite{Lee07, Zhu10, Shopova10, Lu11} single proteins \cite{Armani07}, nucleotides, \cite{Qavi11, Baaske14} and are even used commercially.\cite{genalyte}  Many geometries have been used for bulk refractometric sensing. For example, glass whispering gallery mode resonators such as microspheres \cite{Han05, Lutti08} and toriods \cite{Armani07, Zhu10, Swaim13} exhibit ultra-high quality factors ($Q$s) of $>10^6$ allowing precise readout of optical mode wavelengths, and with tens of nm/RIU sensitivity achieve detection limits of $10^{-7}$ refractive index units (RIU).\cite{Han05}  Glass WGMs with a hollow core, dubbed liquid core optical ring resonators (LCORRs), have been shown to achieve gigantic sensitivities of 570 nm/RIU when carefully engineered such that the optical mode sits largely in the liquid core instead of the glass.\cite{Li10}  With $Q$s of $10^5$ these represent the best bulk refractive index sensors in the literature, achieving a limit of detection of 3.8$\times 10^{-8}$ RIU.

LCORRs are remarkably impressive, but for the purposes of integration - such as into lab-on-a-chip devices - it may be more useful to have WGM resonators fabricated on CMOS compatible chips.  The commercially proven silicon-on-insulator platform has been used to fabricate optical resonators in planar geometries, which allows for full integration.  Simple planar WGM geometries such as disk \cite{Boyd01, Sch07} or ring \cite{Iqbal10, Su11, Ciminelli14} resonators have demonstrated sensitivities up to 160 nm/RIU with $Q$s up to $10^5$.  Slot WGM resonators are of significant interest, due to their ability to be engineered such that the optical mode lies mostly within the slot and outside the resonator medium \cite{Barrios07, Claes09, Carlborg10} demonstrating up to 298 nm/RIU\cite{Claes09} but with $Q$s reaching only a couple thousand; photonic crystal resonators utilize photonic bandgaps to highly localize the optical mode \cite{Adams05, Lee07} and have demonstrated 490 nm/RIU sensitivities with similar $Q$s.  

\begin{figure}[b]
\includegraphics[width=8.5cm]{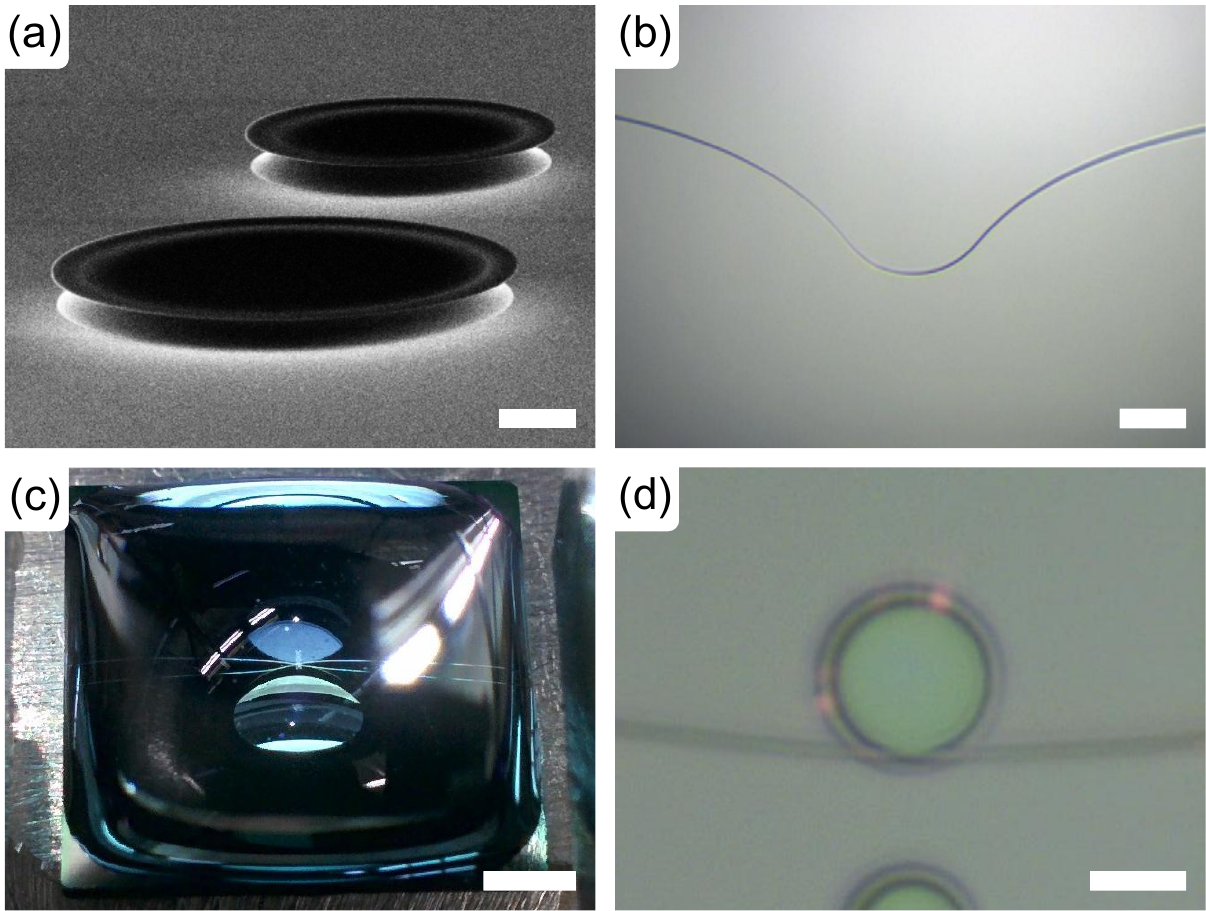}
\caption{\label{fig1}
(a) Scanning electron microscope image of 20 \um and 30 \um microdisks. Scale bar 5 \um. (b) Side view of a dimpled-tapered fiber for visible light used to couple to individual microdisks. Scale bar 100 \um.  (c) A representative $\sim$140 \uL water droplet deposited on a chip of microdisks.  The tapered fiber is visible in the droplet touching the chip.  Scale bar 2 mm.  (d) The dimpled-tapered fiber is used to selectively couple light into a 20 \um microdisk.  On resonance, light in the mode is visible due to surface scattering.  Scale bar 10 \um. 
}
\end{figure}

Here we demonstrate an attractive permutation of an on-chip WGM resonator to be used for refractive index sensing - a thin silicon nitride microdisk resonator.\cite{Barclay06, Baker12}  Si$_3$N$_4$ is a desirable material for optical sensing due to its CMOS compatibility,\cite{Moss13} transparency to visible light, and lower refractive index than silicon resulting in less mode confinement.\cite{Eichenfield07} Si$_3$N$_4$ refractometric sensors have been described previously in optical ring and slot geometries,\cite{Ksendzov05, Barrios07, Carlborg10} and with optimization have achieved sensitivities of 246 nm/RIU and detection limits of $5\times10^{-6}$ RIU.\cite{Carlborg10}  Here we exploit silicon nitride's transparency to 780 nm laser light to enable large portions of the optical field to be in water, negating much of the optical absorption caused by water at longer wavelengths.  Using thin ($< 150$ nm) on-chip Si$_3$N$_4$ microdisks and an under-cut geometry to lower mode confinement, sensitivities of $>200$ nm/RIU and a limit of detection of $\sim$1$\times10^{-6}$ RIU are measured.  This responsiveness results from extending the evanescent field into the aqueous solution, similar to photonic crystal or slot resonators, yet with less stringent fabrication requirements. 

Si$_3$N$_4$ microdisks with diameters of 15, 20, 25 and 30 \um were fabricated and characterized for their bulk index of refraction sensitivity.  Devices were fabricated from silicon wafers with 3 \um of oxide beneath a 150 nm LPCVD deposited stoichiometric Si$_3$N$_4$ film (Rogue Valley Microdevices).  Electron beam lithography was used to pattern an aluminum etch mask, and a SF$_6$ reactive ion etch was used to pattern the nitride. 3 \um of the buried oxide and the Al mask were etched away using a buffered HF solution.  Although electron beam lithography was used to fabricate the current devices,  the large minimum feature size of the microdisks would allow these devices to be fabricated with standard photolithographic processes.

To couple the 780 nm light into and out of the microdisks a dimpled-tapered fiber was used.\cite{Michael07} It was fabricated by tapering an optical fiber (Thorlabs SM600) to $\lesssim$ 1000 nm, the single-mode cutoff diameter for 780 nm light in air,\cite{Hauer14} and then molded to produce a section of the tapered fiber out-of-plane to the rest of the fiber - the dimple - as pictured in \fig{1(b)}.  The fiber is affixed to a pronged mount with the dimple extending downwards towards the chip, which is secured to a 3-axis nanopositioning system, allowing coupling to individual devices fabricated within a planar array.\cite{Hauer14}  An optical camera is used to monitor the coupling procedure and by varying the placement of the fiber, coupling to the microdisk can be tuned. The dimpled-tapered fiber allows operation over a large wavelength range, achieving $>$ 50 \% transmission through the taper over the entire range of our tunable laser (765--781 nm).  Losses in transmission are primarily from non-adiabatic tapering of the optical fiber.  To our knowledge, this is the first time coupling to a planar device with a dimpled-tapered fiber has been demonstrated with visible light in a liquid environment.  

To conduct aqueous experiments, a sample cell is created by depositing $\sim 140$ \uL of deionized water on top of the 10$\times$10 mm wafer, creating a droplet, such as the one demonstrated in \fig{1(c)}, in which the tapered fiber can be submerged.  The system is housed in a closed chamber containing a water reservoir to increase humidity and minimize evaporation of the sample droplet. A micropipette was used to add small volumes of solution to the droplet and induce mixing to homogeneously distribute the solution.  Introducing and removing the pipette tip from the water creates large mechanical oscillations of the droplet which were often enough to move the dimpled fiber a few \um, altering the coupling to the target resonator.  This was minimized by attaching the fiber to the microdisk to provide mechanical stability during the experiment.  This has the disadvantage of lowering quality factors and obscuring detection of the $n > 1$ modes, where $n$ is the radial mode number, in all but the 30 \um disk as well as inducing an additional scattering loss ($\sim$ 50 \%).  Nonetheless, sufficient signal was retained to easily resolve the TE$_\mathrm{n=1}$\cite{JacksonCh8} modes.

\begin{figure}
\includegraphics[width=8.5cm]{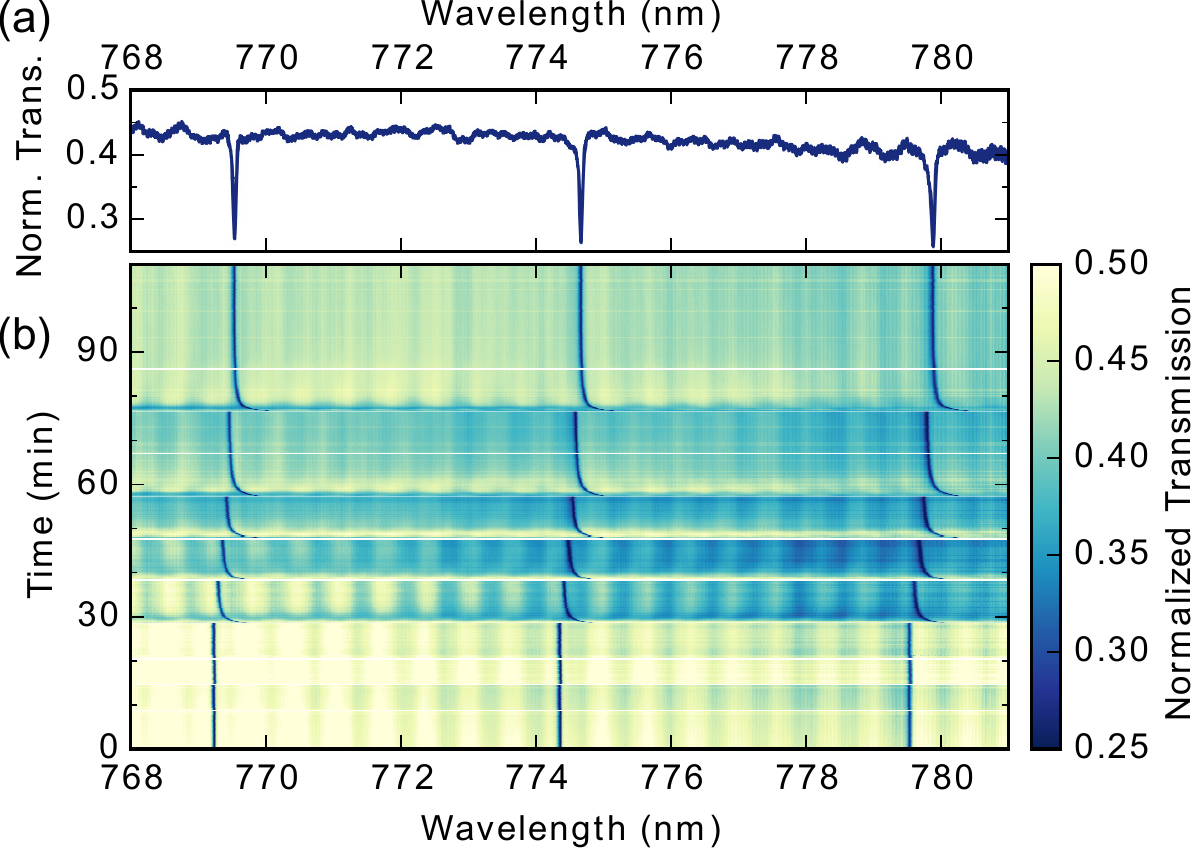}
\caption{\label{fig2}
(a)  The transmission spectra of visible laser light through a dimpled-tapered fiber coupled to a 20 \um diameter Si$_3$N$_4$ resonator.   (b)  By repeatedly scanning a tunable diode laser over its wavelength range, the time dependance of three TE$_{\mathrm{n}=1}$ modes can be tracked as LiCl is added to the environment.  Transmission spectra are normalized to uncoupled fiber transmission.  Events, described in the text, are indicated by white lines.  Spectrum in (a) is taken at 110 minutes.
}
\end{figure}

Once coupled to a microdisk, the transmission spectra of light from a tunable diode laser (NewFocus TLB-6712) can be measured to determine the wavelengths and quality factors of the coupled whispering gallery modes.  Before reaching the disk, the light is attenuated to $\lesssim$ 1 mW to ensure linear behavior of the optical resonances and the polarization is controlled with a three-paddle polarization controller to optimize coupling to TE modes.  Wavelengths are calibrated to the internal wavelength reference of the tunable laser, outputted as a voltage and collected synchronously with the fiber transmission.  A wavelength scan while coupled to a 20 \um diameter microdisk is shown in \fig{2(a)}.  By automating repeated scanning of the laser, time resolved spectroscopy of the optical disk can be performed, allowing the wavelength of multiple whispering gallery modes to be simultaneously tracked, as visualized in \fig{2(b)}.

\begin{figure}
\includegraphics[width=8.5cm]{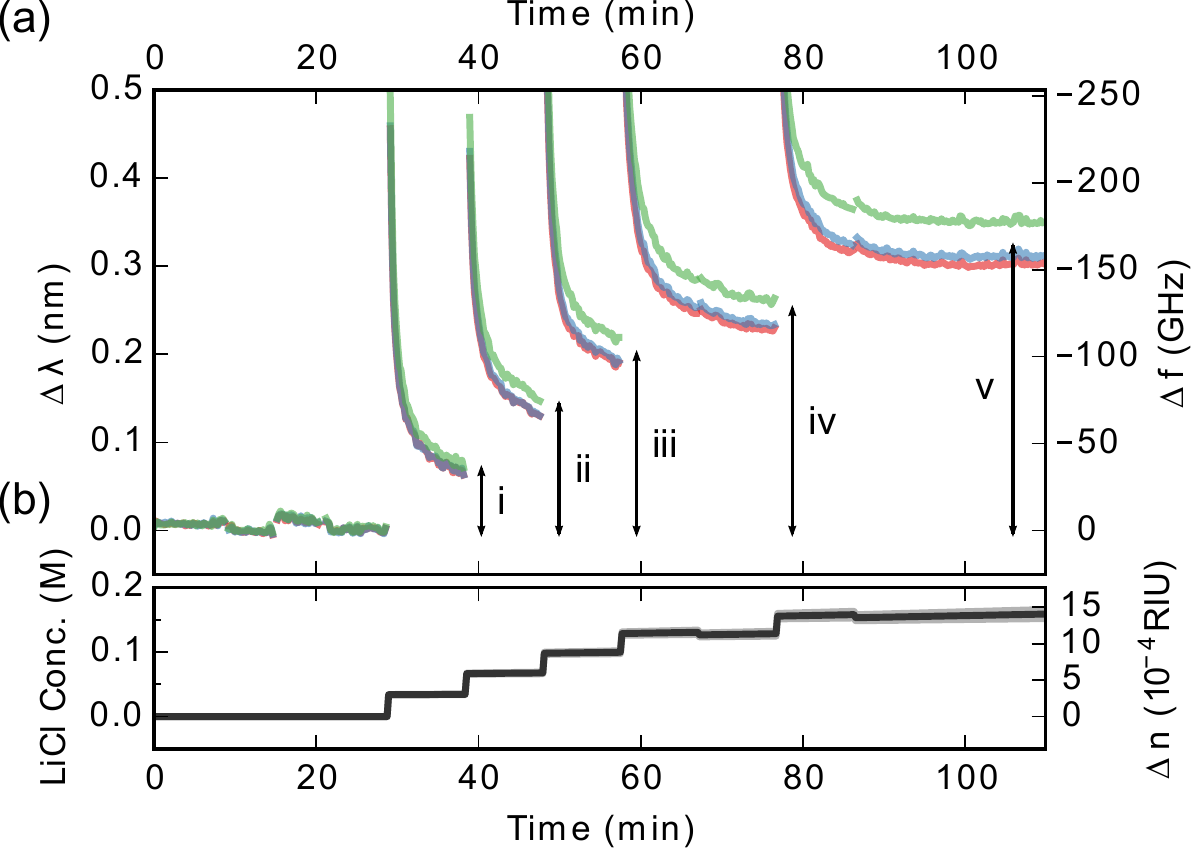}
\caption{\label{fig3} 
(a) Fitting the transmission data in Fig.~2(b) to Lorentzian line-shapes provides readout of wavelength (or frequency) shifts of the whispering gallery modes.   The wavelength shifts marked i through v correspond to bulk refractive index sensitivities of 227, 238, 226, 221, and 229 nm/RIU respectively.  (b)  Small amounts of concentrated Li salt solutions are periodically added to the sample cell increasing the LiCl concentration.  These correspond to refractive index shifts above pure water (1.330 RIU) of 0.00886 RIU/M.  Gray lines indicate concentration error limits, chiefly due to uncertainty in the water evaporation rate.
}
\end{figure}

To measure the bulk refractive index sensitivity of the microdisks, LiCl solutions with concentrations of 1 mol/L (1 M) were added to the sample cell in 5 \uL volumes increasing the refractive index from pure water (1.330) linearly proportional to salt concentration with a slope of 0.00886 RIU/M.\cite{CRCHandbook}  This caused wavelength shifts of the optical modes that, by automated fitting of Lorentzians to the resonances, provide quantitative readout of the wavelength change of each mode. Extracted wavelength shifts for the run in \fig{2(b)} are plotted in \fig{3(a)}.  At 9, 15, 67 and 87 minutes,  5 \uL of deionized water was added and mixed to the droplet, and at 21 minutes the droplet is mixed without adding or removing water.  During these events the wavelength of each mode remained relatively unchanged.  At 29, 39, 58, and 77 minutes 5 \uL of 1 M LiCl was added to the droplets, causing large positive shifts of mode wavelengths.  Interestingly, these events display transient behavior due to diffusion of the ions inside the droplet.  The known times for Li$^+$ and Cl$^-$ ions to diffuse a root-mean-square distance of 1 mm is 4 and 8 minutes respectively - similar to the time scales observed.\cite{CRCHandbook}

By tracking the volumes of deionized water and LiCl solution added to the sample cell, the concentration - and therefore the index of refraction of the environment - was determined.  Uncertainty in the rate of evaporation of water from the droplet gave uncertainties in LiCl concentration as indicated in \fig{3(b)}.  Knowing the wavelength shift and index of refraction of the water allowed the bulk refractive index sensitivities of the whispering gallery modes to be determined and are plotted in \fig{4(a)}, with errors representing standard deviations coming from a combination of variance between addition events (\textit{e.g.}~\fig{3(a) i--v}), and wavelength and concentration uncertainties.

\begin{figure}
\includegraphics[width=8.5cm]{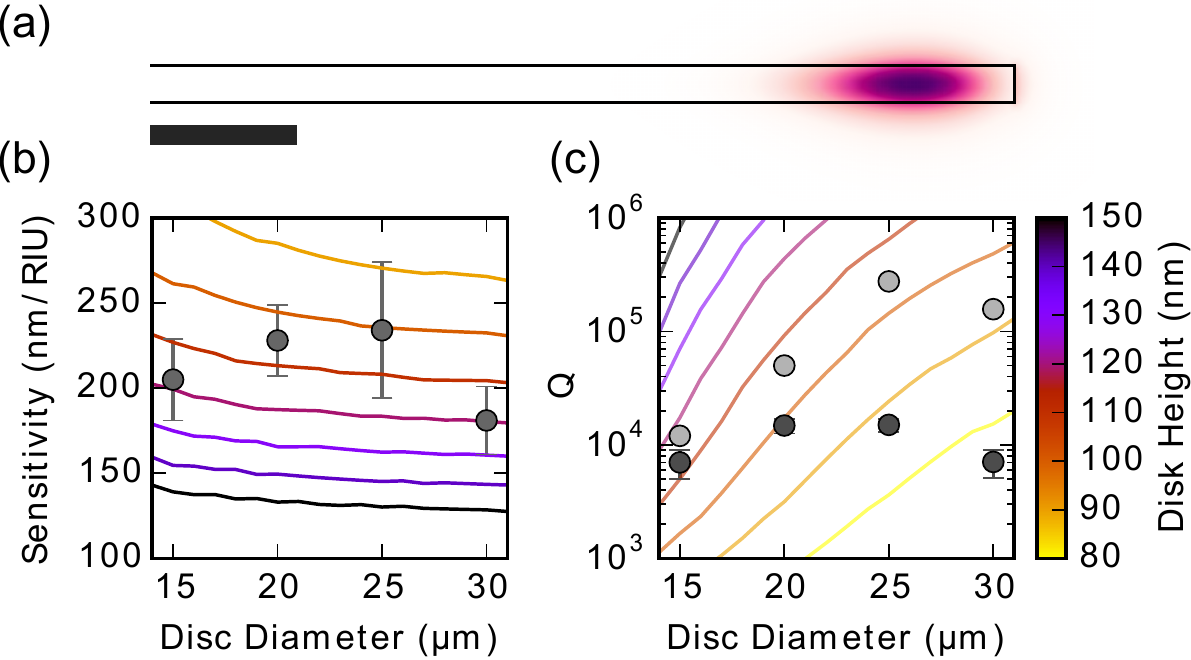}
\caption{\label{fig4} 
(a) The TE$_\mathrm{n=1, m=121}$ mode shape of a 20 \um diameter, 130 nm thick, resonator determined through FEM simulations.  Darker colors indicate larger magnitude electric field.  Scale bar 1 \um.  (b)  Measured sensitivities plotted on-top of simulated sensitivities for a range of disk thicknesses.  (c)  Measured loaded (dark) and unloaded (light) quality factors ($Q$s) plotted on top of simulated radiative loss quality factors.
}
\end{figure}

Axisymmetric simulations of the whispering gallery modes \cite{Oxborrow07} were performed to determine the theoretical refractive index sensitivity of the whispering gallery modes,  as well as the mode quality factors due to radiative losses.  Measured sensitivities for the disks were larger than expected for 150 nm thick disks, \fig{4(b)}, however thinning of the disks during fabrication may explain the enhanced sensitivities.  A Si$_3$N$_4$ etch rate of only 0.3 nm/min would result in 120 nm thick discs.  As expected, observed $Q$s are less than simulated (\fig{4(c)}), as simulations neglect most loss mechanisms.  With the tapered fiber attached to the microdisks, $Q$s of $\sim10^4$ are observed, however hovering the tapered fiber away from the disks reduces fiber-induced losses and reveals unloaded $Q$s of $> 10^5$ for the 25 and 30 \um diameter disks, \fig{4(c)}, as well as mode-splitting of a few pm - which could be used as an additional sensing mechanism.\cite{Zhu10}  Simulations also predict large radiative losses for TM modes, incompatible with the measured $Q$s and therefore provide mode identification as TE, visualized in \fig{4(a)}.

Evaluating the wavelength stability of the microdisks allows us to estimate the refractive index limit of detection of the micodisks.  Taking repeated wavelength scans provides a direct method to estimate the uncertainty in mode wavelength by computing the variance between multiple scans. Further,  averaging successive scans with a low-pass filter provides a method to reduce the uncertainty in wavelength measurement by removing the high-frequency stochastic error in each wavelength sweep.  
This can be improved upon by using the piezo scan functionality of the tunable laser, sacrificing scan range, but increasing wavelength repeatability and allowing calibration with an external wavelength meter.  Scanning a single mode of a 20 \um disk, as shown in \fig{5}, allows wavelengths to be determined with a standard deviation of 0.1 pm over 20 minutes with a filter time constant of only 30 s.  With a 20 \um disk sensitivity of $230 \pm 20$ nm/RIU, this corresponds to a three-standard-deviation\cite{White08} detection limit of $(1.3 \pm 0.1)\times10^{-6}$ RIU, or an LiCl concentration difference of $(1.5 \pm 0.1)\times10^{-4}$ M.

Further work is required before these can be realized as useful sensors.  In particular, we have not addressed specificity towards a particular molecule.  Functionalization of the nitride surface \cite{Banuls10} may provide a solution,  but it is unknown how it will affect optical $Q$s or how the bulk index sensitivity will translate into attached molecule sensitivity.  Nonetheless, extension of aqueous sensing into the visible regime, through the use of thin silicon nitride microdisks and visible-light dimpled-tapered fibers, may allow for easing of the technical requirements in sensing applications, such as the use of cheap diodes and spectrometers for visible wavelengths.

Future improvements will focus on fluid handling.  Currently, non-homogenous mixing of solute, long timescales for mixing, and solvent evaporation create uncertainty in solute concentration.  Additionally, mechanical disturbances of the pipette breaking the water surface may contribute to uncertainty in the mode wavelength.  Therefore integration into a fluid-handling system, such as by using microfluidic devices,\cite{Carlborg10} will be beneficial.

\begin{figure}
\includegraphics[width=8.5cm]{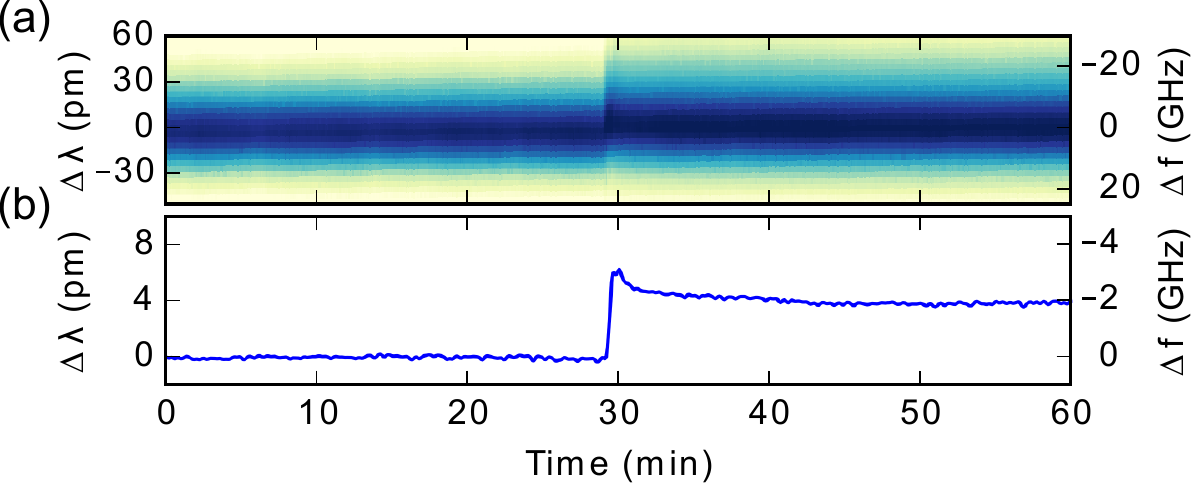}
\caption{\label{fig5} 
(a) Using the piezo scan functionality of the laser allows for better spectral resolution at the expense of scanning range.  (b)   After subtraction of a constant background slope, the wavelength of a mode can be tracked with a standard deviation of 0.1 pm.  At 30 minutes LiCl was added to the sample cell resulting in a $1.5\times10^{-5}$ RIU change, with large signal to noise.
}
\end{figure}

The thin Si$_3$N$_4$ microdisks we have presented are an attractive option for future whispering gallery mode sensors.  The planar configuration allows for mass fabrication and the possibility of integration with lab-on-a-chip technologies.  Sensitivities of $> 200$ nm/RIU are observed, comparable with previously described slot resonators but with lessened fabrication requirements, while maintaining loaded $Q$s of $> 10^4$.  

\begin{acknowledgments}
The authors would like to thank Professor J.M. Gibbs-Davis for helpful discussions and supplying LiCl salt, and A. Darlington for help preparing the solutions.  We also thank A.J.R. MacDonald and B.D. Hauer for mentoring B.C. Lewis under the support of WISEST.  This work was supported by the University of Alberta, Faculty of Science, the Natural Sciences and Engineering Research Council of Canada, Alberta Innovates Technology Futures, and the Canadian Foundation for Innovation.
\end{acknowledgments}

\bibliography{water_wgm}

\end{document}